\documentclass[12pt]{article}
\usepackage[left=2cm, right=2cm, bottom=2cm ,top = 2cm, footskip=1cm]{geometry}
\usepackage[font=small,labelfont=bf]{caption}

\usepackage{setspace}
\usepackage{lineno}

\let\counterwithin\relax
\usepackage{chngcntr}

\usepackage{hyperref}

\usepackage[authoryear,sort&compress]{natbib}
\setcitestyle{open={(},close={)}}

\usepackage{graphicx}
\usepackage{amsmath}
\usepackage{natbib}
\PassOptionsToPackage{hyphens}{url}
\usepackage{hyperref}

\usepackage[section]{placeins}

\title{Multiscale cortical morphometry reveals pronounced regional and scale-dependent variations across the lifespan}
\author{Karoline Leiberg$^{1}$, Timo Blattner$^{2}$, Bethany Little$^{1}$, \\ Victor B.B. Mello$^{2}$, Fernanda H.P. de Moraes$^{3,4}$, Christian Rummel$^{2}$, \\ Peter N. Taylor$^{1,5,6}$, Bruno Mota$^{3}$, and  Yujiang Wang$^{1,5,6*}$}

\begin{document}
\sloppy

\maketitle

\begin{enumerate}
\item{CNNP Lab (www.cnnp-lab.com), Interdisciplinary Computing and Complex BioSystems Group, School of Computing, Newcastle University, Newcastle upon Tyne, United Kingdom}
\item{Support Center for Advanced Neuroimaging (SCAN), University Institute of Diagnostic and Interventional Neuroradiology, University of Bern, Inselspital, Bern University Hospital, Bern, Switzerland}
\item{metaBIO Lab, Instituto de Física, Universidade Federal do Rio de Janeiro (UFRJ), Rio de Janeiro, Brazil}
\item{Brain Connectivity Unit, D’Or Institute of Research and Education (IDOR), Rio de Janeiro, Brazil}
\item{Faculty of Medical Sciences, Newcastle University, Newcastle upon Tyne, United Kingdom}
\item{UCL Queen Square Institute of Neurology, Queen Square, London, United Kingdom}
\end{enumerate}

\begin{center}
* Yujiang.Wang@newcastle.ac.uk    
\end{center}

\newpage
\begin{abstract}

Motivation: Characterising the changes in cortical morphology across the lifespan is fundamental for a range of research and clinical applications. Most studies to date have found a monotonic decrease in commonly used morphometrics, such as cortical thickness and volume, across the entire brain with increasing age. Any regional variations reported are subtle changes in the rate of decrease. However, these descriptions of morphological changes have been limited to a single length scale. Here, we delineate the morphological changes associated with the healthy lifespan in multiscale morphometrics.

Methods: We applied multiscale morphometric analysis to structural MRI from subjects aged 6-88~years from NKI (n=833) and CamCAN (n=641). These multiscale morphometrics were obtained at both the cortical hemisphere and lobe level. 

Results: On the level of whole cortical hemispheres, lifespan trajectories show diverging and even opposing trends at different spatial scales, in contrast to the monotonic decreases of volume and thickness described so far. Importantly, larger scales displayed most dramatic changes across the lifespan (up to 60\%). More pronounced lobal differences in lifespan trajectories also became apparent in scales over 0.7~mm. In a proof-of-principle application in brain age prediction, we also demonstrate added information contributed by multiscale morphometrics.

Conclusion: Our study provides a comprehensive multiscale description of lifespan effects on cortical morphology in an age range from 6-88~years. In future, this can form the foundations for a normative model to compare individuals or cohorts, hence identifying multiscale morphological abnormalities. Our results reveal the complementary information contained in different spatial scales, suggesting that morphometrics should not be considered on a single scale, but as functions of length scale.


\end{abstract}

\newpage
\section{Introduction}

Cortical morphology undergoes significant changes across the lifespan, noticeable even visually. Models of such effects are important for understanding the biological processes which underpin the lifespan, as well as for clinical applications to identify morphological abnormalities.
Previous work describing population trends associated with healthy ageing was predominantly performed in the well-established morphological metrics of cortical thickness, surface area, and volume, and studies usually consider each metric individually. Volume decreases with age have been reported mainly in frontal, but also in parietal and temporal regions \citep{Bethlehem2022, Resnick2003, Storsve2014, Good2001, Salat2004a}. Grey matter volume loss is around 2.4~cm$^3$ per year in adults, with a faster decrease seen in advanced age \citep{Resnick2003}. Regional differences comprise mostly of subtle variations in the monotonic decrease; e.g. a faster decrease is seen in earlier decades or advancing age \citep{Storsve2014, Frangou2021}. Decreases of cortical thickness with age have been described frequently \citep{Bethlehem2022, Fjell2010, Salat2004a, Frangou2021, Storsve2014}, and annual thinning is around 0.35\% across the cortex \citep{Storsve2014}, with maximum thinning estimated around 0.07~mm per decade \citep{Salat2004a}. Regionally, the frontal cortex is generally reported to see more thinning than other areas \citep{Fjell2010, Thambisetty2010, Salat2004a, Storsve2014}.
Cortical surface area has also been shown to decrease with age \citep{Bethlehem2022, Salat2004a, Storsve2014}, with a mean annual change of 0.19\% and the biggest changes seen in the medial temporal, occipital, and posterior cingulate cortices \citep{Storsve2014}.
Overall, with the exception of the first decade, a monotonic reduction of average thickness, volume, and surface area is seen, with little regional differences in these trends.

Prior studies describing lifespan effects on cortical morphology combined information from all length scales down to the  ``native scale'' of the brain surface reconstruction. Such analyses thus lacked a delineation between morphological changes happening at specific scales. We therefore suggest that a multiscale analysis is necessary to capture the contrasts between changes at the level of small cortical features (single gyrus), and the more extensive effects happening at larger scales (lobes and hemisphere level).

To enable such multiscale analyses, we have recently  proposed a re-conceptualisation of cortical morphology \citep{wang_neuro-evolutionary_2023}, in which distributed shape information is quantified across multiple length scales. Rather than measuring a single value of e.g. volume of the cortex, we calculate how morphological features of different sizes separately contribute to the total value. By deleting features smaller than a desired cut-off length scale, we are able to re-render any given cortex to retain only those larger features. By varying the cut-off scale, our focus can then range from the smallest sulci and gyri to entire cortical hemisphere. 


In this study, we characterise morphological changes of the healthy lifespan past the first decade in a multiscale analysis. We begin by describing whole-brain changes, followed by regional differences in individual lobes. Finally, in a proof-of-principle study, we estimate brain age from a single scale \textit{vs.} multiple scales. This brain age model is not optimised for performance or designed to compete with existing models, but to illustrate the added value of multiscale morphometry. Taken together, we provide a comprehensive model of the lifespan, which could serve as a normative model in the future for both individuals and cohorts of patients.

\section{Methods}

\subsection{Data and preprocessing}

To study lifespan effects on cortical morphology, we used T1~weighted MRI of healthy subjects from two large public datasets, the Nathan Kline Institute Rockland Sample (NKI) \citep{Nooner2012}, and the Cambridge Centre for Ageing and Neuroscience (CamCAN) dataset. Both datasets were acquired on a 3T~Siemens TIM Trio scanner with a 1~mm isotropic voxel size (for more details see \url{https://fcon_1000.projects.nitrc.org/indi/enhanced/mri_protocol.html} for NKI, \citep{Shafto2014, Taylor2017} for CamCAN).

\begin{figure}[h!]
\centering
\includegraphics[scale=0.8]{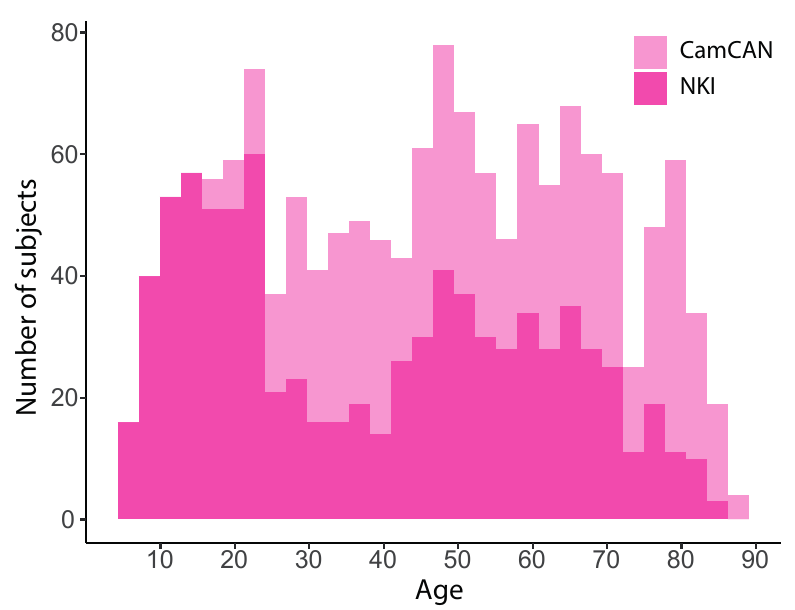}
\caption{\textbf{ Age distribution of subjects used for this study.} Total count of subjects available by age, stacked across datasets and coloured by dataset. } \label{FigAge}
\end{figure}

Of the NKI dataset, 96 images were rejected due to insufficient image quality and motion artifacts, either because they failed processing, or after processing they were identified as outliers and visual inspection attributed this to the quality of the raw image. This left 833 subjects from this dataset (325/508~m/f). 641 subjects of the CamCAN dataset completed processing (327/314~m/f). The age distribution of the data used can be seen in Fig.~\ref{FigAge}.

The MRI were preprocessed with the FreeSurfer~6.0 recon-all pipeline, from which we obtained pial and white matter surface reconstructions, labelled with the Desikan-Killiany parcellation. The surfaces were visually quality checked and corrected where necessary.

All analyses were performed on anonymised data, which were acquired previously as part of other studies/consortia, with ethical approval from Newcastle University (reference: 22/SC/0016).

\subsection{Coarse-graining and regional computation of multiscale morphometrics}

\begin{figure}[h!]
\centering
\includegraphics[scale=0.8]{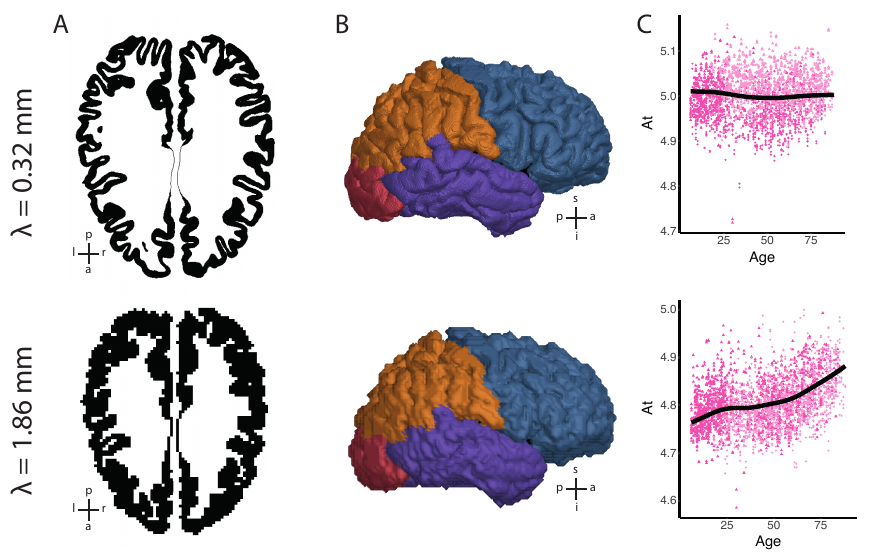}
\caption{\textbf{Computation of lifespan trajectories in multiscale morphometrics in cortical regions.} Algorithm is shown for two example scales, $\lambda =$~0.32~mm (top row) and $\lambda =$~1.86~mm (bottom row). The algorithm was repeated for scales between 0.32~mm and 3~mm. \textbf{A)}~Coarse-grained grey matter and white matter volumes. \textbf{B)}~Reconstructed grey matter surfaces, with lobes labelled using the nearest point on the original FreeSurfer reconstruction. \textbf{C)}~Harmonisation across sites and inference of lifespan trajectories of pial surface area~$A_t$ with gamlss models. } \label{FigMethod}
\end{figure}

We performed a coarse-graining of the cortical surfaces following the algorithm as described in \citep{wang_neuro-evolutionary_2023}. Briefly, the detailed pial and white matter surfaces that were reconstructed by FreeSurfer were converted to a volume, by filling them with voxels of side length~$\lambda$. The voxels were labelled as being within the grey matter or the white matter (Fig.~\ref{FigMethod}~A). This provides a segmentation rendered at scale~$\lambda$, i.e. grey/white matter volumes which consist of voxels of side length~$\lambda$ and are therefore coarser than the original surfaces. We then reconstructed surfaces over the coarse-grained volumes using MATLAB function isosurface (Fig.~\ref{FigMethod}~B), to obtain coarse-grained surfaces, in which folding features smaller than~$\lambda$ have been removed. These surfaces are different to a downsampled rendering of the FreeSurfer surfaces, in that small folding details was intentionally removed, rather than attempting to describe the same shape using less information. We used the FreeSurfer localGI pipeline to obtain smooth (exposed) surfaces of the coarse-grained volumes with a sphere of 15~mm diameter used for the closing operation. We also obtained a convex hull over the coarse-grained pial surface.
 
For the hemisphere analysis, we computed the total surface area $A_t(\lambda)$ of the coarse-grained pial surface, the surface area of the convex hull $A_e(\lambda)$, and estimated the average cortical thickness $T(\lambda)$ as the ratio of the volume of the grey matter (obtained from the coarse-grained mask in Fig.~\ref{FigMethod}~A) and $A_t(\lambda)$.

We followed the FreeSurfer assignment of Desikan-Killiany regions into insula, frontal, parietal, temporal, and occipital lobes (\url{https://surfer.nmr.mgh.harvard.edu/fswiki/CorticalParcellation}). We labelled vertices on the coarse-grained pial, white matter, and exposed surfaces by the label of the closest vertex on the original pial/white matter surface (Fig.~\ref{FigMethod}~B). From this, we computed the lobe-wise pial surface area $A_t(\lambda)$ and exposed area $A_e(\lambda)$. We estimated the average thickness $T(\lambda)$ of each lobe as the average minimum distance from each vertex of that lobe on the pial surface to the white matter surface. The pial surface area allocated to the insula was divided into the frontal, parietal, and temporal lobe according to the relative surface area of those lobes. Likewise, we also included the insula measurement of $T(\lambda)$ into the averages of these three lobes according to their relative surface areas.

We repeated this algorithm for scales~$\lambda$ ranging from 0.32~mm to 3.02~mm (steps starting at $10^{-0.5}$, then increase by multiplying by $10^{0.07}$, e.g. second step $10^{0.07}*10^{-0.5} = 0.37$). We chose this range as it covers surface reconstructions from a very close approximation of the original FreeSurfer surface ($\lambda =$~0.32~mm), to quite a coarse surface, at which differences between individual brains are still visible ($\lambda =$~3~mm). Nonlinear sampling, with more scales sampled at the lower end, ensured we captured the faster changes happening to the surface reconstructions at small scales, whereas at larger scales the surfaces change more gradually. An analysis of even larger scales can be found in the supplementary material (see~\ref{supplMoreScales}).

Note that the scales used for coarse-graining do not directly correspond to acquisition resolutions, for example a surface coarse-grained at $\lambda =$~1~mm is not equivalent to the FreeSurfer surface reconstruction of a 1~mm isotropic image, but will have lost some of the folding detail. In fact, no matter how small the coarse-graining scale, it is impossible to fully reconstruct the original surface with this method, since surfaces between touching gyri walls will not be recovered. At the lower end of the range of scales, small changes in $\lambda$ have a larger effect on the resulting surface reconstruction. Because of this, we sample more densely from the smaller scales, to capture and describe multiscale effects thoroughly.

We continued the analysis with the morphometrics average thickness~$T(\lambda)$ and pial surface area~$A_t(\lambda)$, computed at hemisphere level and for each lobe at a range of spatial scales~$\lambda$.

\subsection{Analysis of lifespan effects on morphometry in cortical hemispheres and lobes}

To extract the lifespan effect on different morphometrics from the two datasets, we used generalised additive models for location, scale, and shape (GAMLSS) from the gamlss R package (\url{https://cran.r-project.org/web/packages/gamlss}). We included the acquisition site as a random variable affecting the mean and standard deviation of the distribution, to account for scanner and scanning protocol effects and harmonise the data between sites. Since we were not primarily interested in sex differences, we accounted for these with random variables as well, modelling the effect on mean, standard deviation, and skew, but producing a single overall trajectory for both sexes (Fig.~\ref{FigMethod}~C). The age effect was modelled on all four distribution parameters with smooth terms using a P-splines. Model formulas can be found in the supplementary \ref{supplFormulas}. The model was fit with a mixed algorithm (``Rigby and Stasinopoulos'', then ``Cole and Green''). We fit these models for hemispheres and cortical lobes, and for each scale, which allowed us to infer hemisphere and regional scale-dependent ageging trajectories, whilst accounting for covariates and harmonising between sites.

\subsection{Multiscale metrics as brain age predictors}

To illustrate the complementary information contained at different spatial scales, we performed a brain age estimation using only a single morphometric computed for cortical hemispheres. To correlate cortical shape to ageing, one needs to extract morphometric variables from the images. The choice of the variables to use is thus, implicitly, a choice of which morphological information to retain, and which to discard.

The most general description of a cortex is that of a thin sheet of gray matter, folded in complex patterns around white matter. Total area ($A_t$) alone cannot tell apart, say, a smooth cortex from a tightly crumpled one; and it is known that ageing affects big and small sulci and gyri differently. To capture that distinction, one needs some indication of how much of this area can be attributed to folded features of different sizes. By applying the coarse-graining algorithm, we geometrically removed sulci and gyri smaller than some threshold scale, and measured the area of resulting coarse-grained cortex. Progressively increasing the threshold allowed us to quantify the contribution of the various fold sizes to the aggregate total area.

We used the mgcv R package to fit generalised additive models (GAM), with fixed effects of sex and site, and smooth terms of the predictor $A_t$ to estimate a subject's age. We fit three models to the data: One using $A_t$ computed at scale 0.32~mm, a second using $A_t$ computed at scale 1.86~mm, and a third using $A_t$ at both scales together. The specific scales were selected for the model as the smallest, closest to ``native'' scale usually be used for quantitative morphometry, and a much larger scale, which previously resulted in the biggest effect size discriminating between young and older groups of subjects \citep{wang_neuro-evolutionary_2023}. The three models were thus:

$$\textrm{Model 1: } Age \sim s(A_{t} (\lambda = 0.32mm)) + Sex + Site$$
$$\textrm{Model 2: } Age \sim s(A_{t} (\lambda = 1.86mm)) + Sex + Site$$
$$\textrm{Model 3: } Age \sim s(A_{t} (\lambda = 0.32mm)) + s(A_{t} (\lambda = 1.86mm)) + Sex + Site.$$

Here, $s(A_t (\lambda))$ is a smooth term (with a basis of thin plate regression splines) of the measure $A_t$ computed at scale~$\lambda$, and $Sex$ and $Site$ are fixed effects. This is a very simple model, which could easily be improved by adding more metrics, scales, and regional information. We intentionally decided to keep the model simple to demonstrate the effect of adding multiscale information.

We ran 1000 bootstraps for each model, computing the root mean squared error (RMSE) each time, to get a distribution of RMSE for each model as a measure of model fit.

\section{Results}

\subsection{Lifespan trends measured at different scales have opposing trajectories within a single metric}

\begin{figure}[h!]
\centering
\includegraphics[scale=0.9]{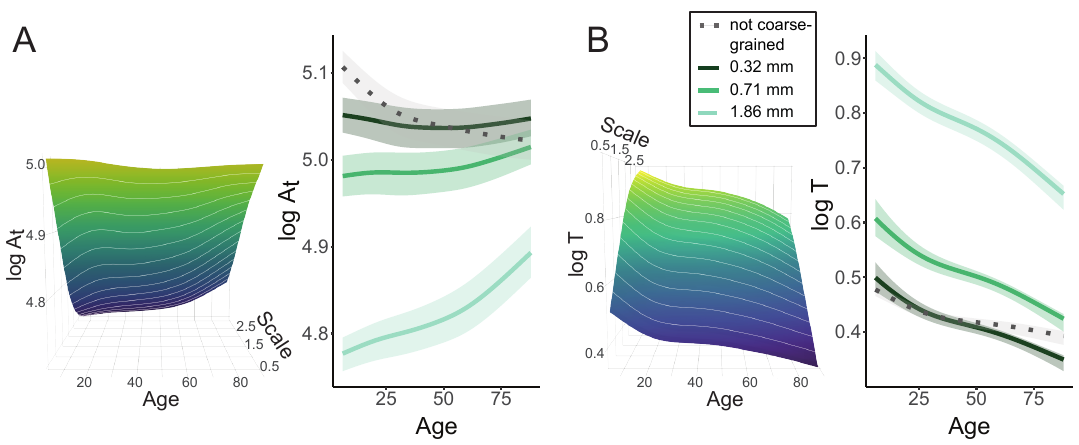}
\caption{\textbf{Lifespan effects on cortical hemispheres measured in multiscale morphometrics.} \textbf{A)}~Pial surface area log10($A_t$/mm$^2$). \textbf{B)}~Average cortical thickness log10($T$/mm). Sheets (left) show trajectories as functions of scales between 0.32~mm and 3~mm. Line graphs show trajectories without coarse-graining (``native scale'', dashed line) and for three scales 0.32~mm, 0.71~mm, and 1.86~mm, where lighter colour indicates a larger scale used for the coarse-graining procedure. Shaded bands show the interquartile range of the distribution. The relative ordering of values of $T$ for the different scales results from a cortex that becomes both thicker and less gyrified as the cut-off scale increases.} \label{FigHemi}
\end{figure}

The GAMLSS modelling produced lifespan trajectories in each scale and morphometric. In  Fig.~\ref{FigHemi}, we show trajectories for all scales as 3D ``sheets'', and cross-sectional trajectories for three representative scales (0.32~mm, 0.71~mm, 1.86~mm) as line graphs. For reference, we also depict the trajectory obtained from the original, not coarse-grained (``native scale'') FreeSurfer surfaces as a dashed line. The three scales shown cover the range from a close approximation of the full, detailed FreeSurfer pial surface (0.32~mm), to a coarse-grained surface with few folding details remaining (1.86~mm). Cross-sectional trajectories of a wider range of scales can be found in the supplementary material~\ref{supplMoreScales}, which provide a thorough multiscale description of the general morphological changes in hemispheres across an age range of 6-88 years. 

Whilst the pial surface area measured at a scale of 0.32~mm remains relatively constant throughout the lifespan (less than 5\% change), we see an increase with age in the larger scales, particularly after 50+~years (Fig.~\ref{FigHemi}~A). For example, at 1.86~mm, we observe a 25\% increase in surface area across the sampled life span.

We further observed a steep decrease of average cortical thickness throughout childhood, and a slower decrease in adulthood visible in all scales (Fig.~\ref{FigHemi}~B). We note, however, that the rate of decrease is most pronounced in larger scales, e.g. at 1.86~mm, we observe a nearly 60\% decrease compared to the 40\% at 0.32~mm across the sampled life span.

In addition to commonly used metrics of cortical thickness and surface area, we also explored lifespan effects in a set of new morphology measures that are statistically independent~\citep{Wang2021}. Supplementary~\ref{supplKS} lays out the rationale for considering independent metrics, their computation, and hemisphere and regional lifespan effects.

\subsection{Regional differences in lifespan effects become apparent in larger scales}

\begin{figure}[h!]
\centering
\includegraphics{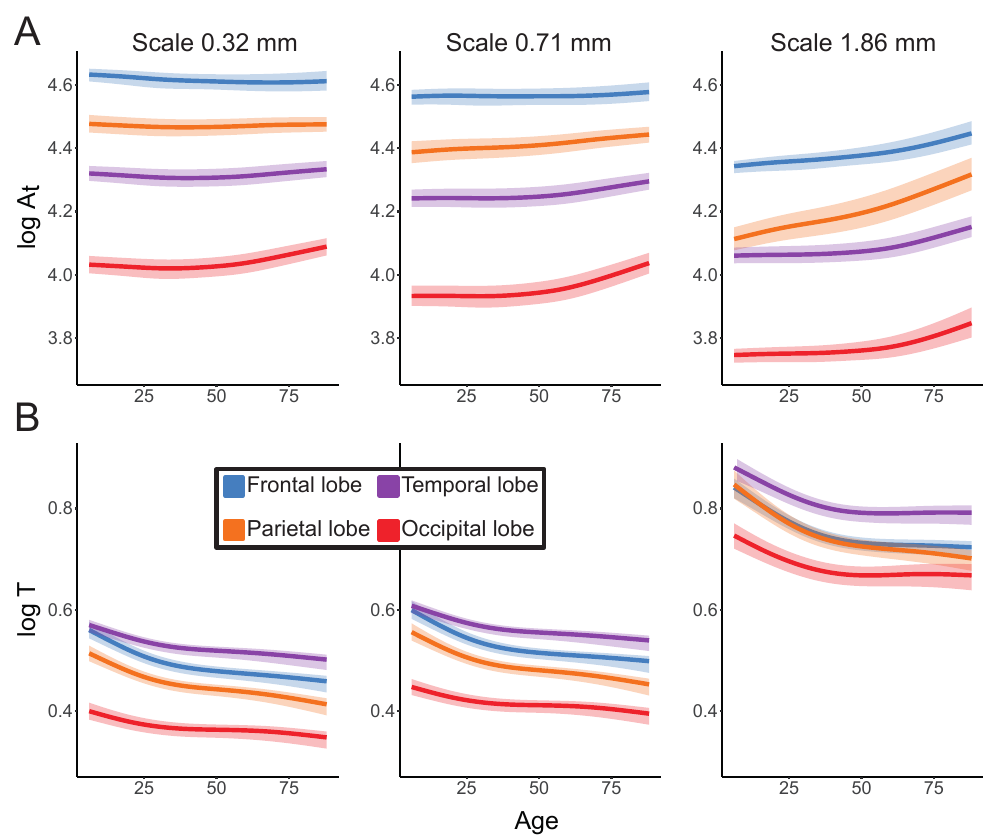}
\caption{\textbf{Lifespan effects in main lobes measured in multiscale metrics.} Columns show trajectories in spatial scales 0.32~mm, 0.71~mm, and 1.86~mm. \textbf{A)}~Pial surface area log10($A_t$/mm$^2)$. \textbf{B)}~Average cortical thickness log10($T$/mm). Colours indicate individual trajectories of cortical lobes. Shaded bands show the interquartile range of the distribution.} \label{FigLobe}
\end{figure}

Next, we investigated lobe-wise lifespan effects across scales. We found that lobe-level changes of pial surface area $A_t$ across the life span were mostly reflected in the coarse scale of 1.86~mm, where most small folding details have been removed. Here, we saw a sharper incline in the parietal lobe across the lifespan (approx 60\%), and a shallower trajectory in the temporal lobe (approx 20\%, see Fig.~\ref{FigLobe}~A). 

In cortical thickness $T$ we found slight regional differences in trajectory in childhood: e.g. at scale 0.32~mm, the cortex thins quicker in the frontal and parietal lobes compare to temporal and occipital lobes.  However, lobal differences during adulthood were minimal, and we gained little additional information across different scales (Fig.~\ref{FigLobe}~B).

\subsection{Brain age estimate from $A_t$ is improved by using morphometrics from multiple scales}

As a proof-of-principle, to demonstrate the added value of multiscale morphometrics in an application, we carried out a simple brain age estimation using only the metric of pial surface area $A_t$. Note this brain age model is not optimised for performance or designed to compete with existing models, but only to illustrate the added value of multiscale morphometry.

\begin{figure}[htp]
\centering
\includegraphics[scale=0.7]{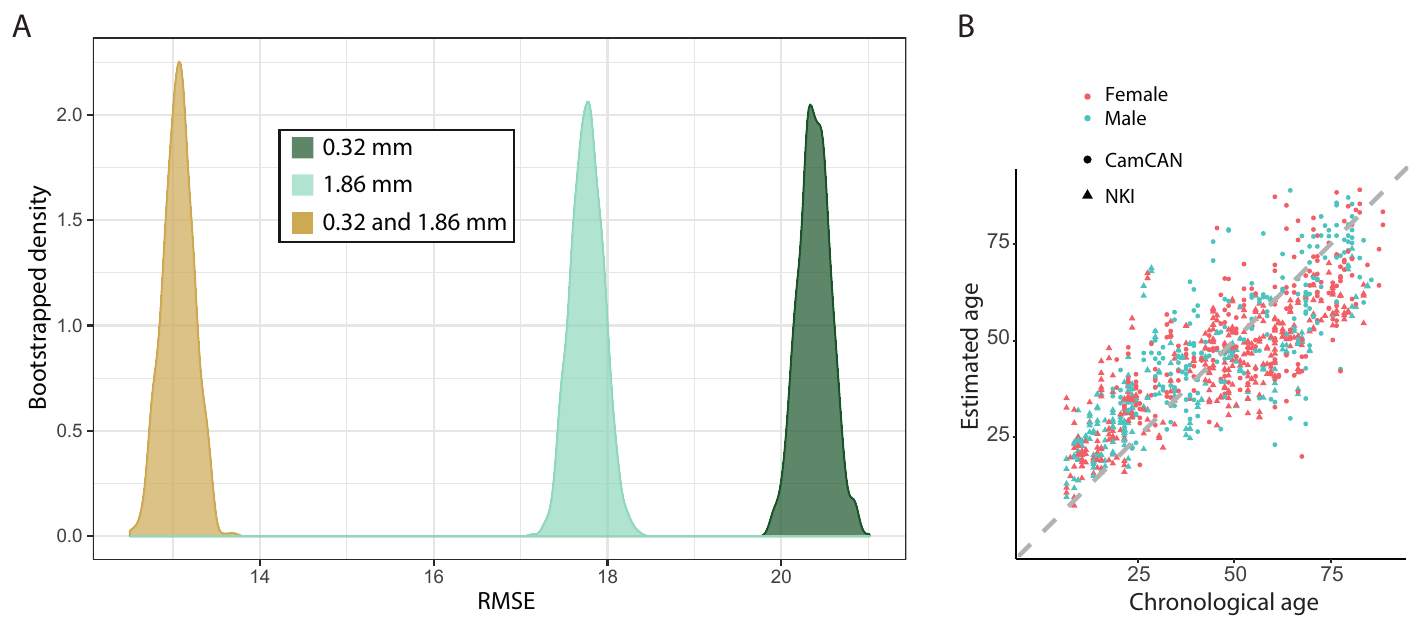}
\caption{\textbf{Brain age estimation using surface area, sex, and site.} \textbf{A)}~Distributions of bootstrapped RMSE obtained from GAM brain age model with effects of age, site, and pial surface area computed at scale 1.89~mm, scale 0.32~mm, and both scales 1.89~mm and 0.32~mm. \textbf{B)}~Fit of model using $A_t$ computed at both scales, showing estimated and actual (chronological) age of subjects. } \label{FigBrainAge}
\end{figure}

We compared model performance when using a single morphometric to using multiple scales (Fig.~\ref{FigBrainAge}). Both of the models using $A_t$ computed at a single scale (0.32~mm and 1.86~mm, respectively) had high RMSEs of over 17~years.wever, using surface area measurements at both scales 0.32~mm \textit{and} 1.86~mm, the model fit was improved to an RMSE of under 14~years, showing that model performance was enhanced by using just a single metric at different scales.

As a side note, a straight line fit through the data in Fig.~\ref{FigBrainAge}~B would have a slope lower than~1, indicating that our model displays a slight bias to the mean age of the dataset (48~years). This ``regression to the mean'' is a known phenomenon in brain age estimation, that can be accounted for by performing an age regression on the estimated values before further analysis \citep{Le2018, Liang2019}. Since we are not using the estimated brain age for further applications here, we omitted this step.

\section{Discussion}

In this study, we analysed lifespan effects on cortical morphology. We quantified changes in a range of morphological metrics across spatial scales, and found divergent trajectories with increasing age at different scales. Furthermore, we also observed unique trajectories in different lobes at larger spatial scales. Our findings enrich current reports in the literature, which have focused only on a single ``native'' scale analysis, and mostly found monotonically decreasing lifespan trajectories. Clearly, different types of morphology information are contained at different spatial scales, as evidenced by the divergent lifespan trajectories. As a proof-of-principle, we also demonstrated that these different types of information can be leveraged to improve predictive performance.

\subsection{Multiscale cortical morphometry}

Here, we used a multiscale approach based on computationally coarse-graining the surface representation of the brain. This allows us to create coarse-grained biologically plausible \citep{wang_neuro-evolutionary_2023} reconstructions of the brain for specific length scales, from which we can compute scale-specific morphometrics.

Such scale-specific morphometrics allow us to disentangle the shape information of features of varying size, since specific processes and pathologies might affect cortical shape at different scales. For example, lesions may show as localised alterations at small scales, whilst dementias might cause atrophy in wide areas and large-length scales. In future applications, this new quantification of morphology could improve methods for diagnosis, whilst also being more precise in delineating diverging morphological trajectories.

Other complementary multiscale descriptions of cortical morphology have also been proposed, such as spectral analysis of cortical geometry \citep{Chen2022}. This approach also characterises cortical morphology across spatial scales, but assumes basis functions, and the interpretation of eigenvalue spectra is thus challenging due to possible harmonics caused by the specific choice of the basis function. Additionally, key differences to our approach include: intrinsic co-variates (such as isometric brain size) are not factored out in a way that respects scaling relationships, and the cortical shape is only analysed for the white matter surface. Nevertheless, we are excited that other researchers have also turned their attention to multiscale morphometrics. We believe that future iterations of this approach will combine multiple methods and open the door to exciting discoveries in morphological analysis of (human) brains.

\subsection{Lifespan trends in cortical hemispheres}

At the smallest scales, our surface reconstruction is a close approximation to the detailed FreeSurfer surface, and indeed the morphological trajectories we observed at a scale of 0.32~mm agree with previous descriptions of lifespan effects \citep{Bethlehem2022, Frangou2021} and our trajectories of metrics computed from the original surfaces. In the hemisphere analysis, the coarse-graining affects average cortical thickness mostly by changing the offset rather than the shape of the trajectory. This indicates that lifespan effects in this metric are largely scale-invariant, meaning cortical thinning affect all scales similarly. However, we found diverging trajectories at different scales in pial surface area. This shows that this metric is able to capture changes in cortical shape uniquely in different scales, due to sulci widening with advancing age, an effect that is mostly captured in larger spatial scales.

\subsection{Regional differences in age trends}

Our multiscale analysis of age trends in cortical lobes revealed some regional differences that were not detectable at the smallest scales alone. We found that pial surface area increased with age in the occipital lobe even in the smallest scales, likely caused by the tight cortical folds here, which accelerate the coarse-graining procedure. We found minimal regional differences in average cortical thickness, but what we did find is in agreement with existing literature. Namely, a difference in offset of trajectories, and steeper and longer slope during development in frontal and parietal cortices \citep{Frangou2021}.

Overall, we found that additional scales contain important, new, nonredundant information, making them able to distinguish regional lifespan effects much better than traditional metrics and computation methods. Further, they are possibly pointing to biological processes in lifespan affecting the shape of cortical regions differently.

\subsection{Utilising multiscale information}

We built a model for brain age estimation using only sex, scanning site, and pial surface area as predictors. The estimates were vastly improved by including surface area measurements from two scales, rather than just a single measurement at one scale. The model's performance does not approach that of current best brain age estimators using morphometrics, which achieve mean absolute errors of under 4~years, for example by combining information from multiple modalities \citep{Cole2020, Rokicki2021, Guan2023}. Whilst our model could quickly be improved by adding more metrics, scales, and regions, we kept it simple, since our goal was to highlight the complementary information that is contained at different spatial scales and the value we can extract from it. It suggests potential for future brain age estimators, that build on this result, creating models that are both accurate and interpretable. Of course, brain age estimation is just an illustrative application area of multiscale morphometry, and our results suggest that analyses in other applications, such as diagnosis or structural abnormality detection, might be improved with a multiscale analysis approach.

\subsection{Normative models of healthy ageing}

Normative models are the mapping of health-related variables to each other, and result in population-level trajectories \citep{Rutherford2023}. Whilst they capture variation in the population, they also allow us to quantify individual variation and deviation from the population. This both enhances group-level inference, but also enables subject-level analysis, for example in MRI analysis finding abnormal/atypical trajectories of structural change \citep{Rutherford2022}. Hence, they can help contrast trajectories of degenerative diseases, show how and when they diverge from healthy ageing, and describe changes in disease cohorts, isolated from ageing.

In our study, we built multiscale models of morphometrics computed at a range of spatial scales. This multiscale approach has the potential to improve sensitivity and specificity of the model, since previous normative models based on multimodal data have shown improved performance due to the complementary information contained in the data \citep{Kumar2023}.

\subsection{Limitations}

Even though our inference of lifespan effects is based on two large data sets, the analysis could be improved by including data from additional sites, making the trajectories more robust and generalisable. In particular, the effects in ages below 18 years were only based on the NKI data. In future work, we will add more developing cohorts so that the analysis is not based on a single dataset for this age range.

Currently, our regional analysis is limited to cortical lobes. We were able to show local lifespan effects in larger scales at this level already, but we will develop methods to further localise the computation of multiscale morphometrics for more precise regional analyses in the future.

\subsection{Conclusion}

Our study describes healthy lifespan effects on multiscale morphometrics. We found that metrics computed at different spatial scales capture distinct aspects of cortical shape, suggesting that there could be different biological mechanisms underlying the effects that uniquely impact individual scales and cortical regions. We also show that larger spatial scales should be leveraged for their additional information in morphometric analyses.

\section{Data and Code availability}

The analysis was carried out on public datasets, see \url{http://rocklandsample.org/} for NKI and \url{https://www.cam-can.org/} for CamCAN.

The coarse-graining of cortical surfaces was performed using code available on github: \url{https://github.com/cnnp-lab/CorticalFoldingAnalysisTools/blob/master/Scales/}.

\section{Author contributions}

KL: Conceptualisation, Investigation, Methodology, Data Curation, Formal analysis, Visualisation, Writing - Original Draft. TM: Methodology, Writing - Review \& Editing. BL: Methodology, Data Curation, Writing - Review \& Editing. VBBM: Methodology, Writing - Review \& Editing. FHPM: Methodology, Writing - Review \& Editing. CR: Funding acquisition, Methodology, Writing - Review \& Editing. PNT: Funding acquisition, Writing - Review \& Editing BM: Funding acquisition, Conceptualisation, Methodology, Writing - Review \& Editing. YW: Conceptualisation, Supervision, Funding acquisition, Methodology, Writing - Review \& Editing.

\section{Funding}

K.L. is supported by the EPSRC Centre for Doctoral Training in Cloud Computing for Big Data (EP/L015358/1). P.N.T. and Y.W. are both supported by UKRI Future Leaders Fellowships (MR/T04294X/1, MR/V026569/1). B.L. and Y.W. are supported by the EPSRC (EP/Y016009/1). C.R. and Y.W. are supported by the Swiss National Science Foundation (SNF, grant 204593, "ScanOMetrics" project). B.M. is supported by Fundação Serrapilheira Institute (grant Serra-1709-16981) and CNPq (PQ 2017 312837/2017-8).

\section{Declaration of Competing Interests}
The authors declare no conflicts of interest.

\section{Acknowledgements}

We thank members of the Computational Neurology, Neuroscience \& Psychiatry Lab (www.cnnp-lab.com) for discussions on the analysis and manuscript.

\newpage
\bibliography{NormativeMelting}
\newpage


\renewcommand{\thefigure}{S\arabic{figure}}
\setcounter{figure}{0}
\counterwithin{figure}{section}
\counterwithin{table}{section}
\renewcommand\thesection{S\arabic{section}}
\setcounter{section}{0}

\section*{Supplementary}

\section{Model formula} \label{supplFormulas}

We modelled the morphological data using flexible sinh-arcsinh (shash) distributions. This distribution accommodates non-normal data modelling, allowing the first four moments to vary as functions of explanatory variables. All metrics were log-transformed prior to statistical analysis. Using gamlss, we modelled the distribution's parameters (moments) as dependent on explanatory variables, specifically sex, age, and scanning site. We fitted this model to each metric separately. The model formulae we used were:

\begin{equation*} \label{eq1}
\begin{split}
\mu & \sim 1 + (1 | sex) + s(age) + (1 | site) \\
\sigma & \sim 1 + (1 | sex) + s(age) + (1 | site) \\
\nu & \sim 1 + (1 | sex) + s(age) \\
\tau & \sim 1 + s(age)
\end{split}
\end{equation*}

\begin{itemize}
    \item The mean ($\mu$) depends on sex (random effect), site (random effect), and a smooth function of age.
    \item The standard deviation ($\sigma$) depends on sex (random effect), site (random effect), and a smooth function of age.
    \item The skew ($\nu$) depends on sex (random effect) and a smooth function of age.
    \item The kurtosis ($\tau$) depends on a smooth function of age.
\end{itemize}

\section{Independent morphometrics} \label{supplKS}

Beyond the commonly used metrics of thickness and surface area, we also applied a set of independent morphometrics described previously \citep{Wang2021}. Briefly, a scaling law of cortical folding has been proposed \citep{Mota2015} and empirically validated \citep{Wang2016, Wang2019, Leiberg2021}. The scaling law allows the definition of a set of new morphometrics, that are independent of each other and have physically meaningful interpretations \citep{Wang2021}. Healthy lifespan trajectories have also been inferred in these new morphometrics \citep{DeMoraes2022}. We performed morphological analysis in these new metrics alongside the commonly-used metrics of thickness and surface area in our analyses.

\subsection{Computation}

Specifically, we computed two dimensionless, independent morphometrics $K(\lambda)$ and $S(\lambda)$ as linear combinations of the logarithms of $T(\lambda)$, $A_t(\lambda)$, and $A_e(\lambda)$. We interpret these terms as the summary expressions of, respectively, a near-invariant tension term associated with the conserved dynamics of axonal elongation in a white matter surrounded by a self-avoiding grey matter; and a varying shape term that summarizes the morphological complexity of a gyrified structure. These metrics account for the covariance in $A_t$, $A_e$, and $T$, and we have found them to be useful for describing cortical shape, being better at differentiating morphological processes and capturing more subtle changes in cortical shape \citep{Wang2021, Leiberg2021}.

These metrics were initially well-defined for complete cortical hemispheres. To derive these metrics for individual cortical lobes, we computed the integrated Gaussian curvature over the convex hull for each lobe, and used it to correct the surface area measures to what their value would be for an entire cortical hemisphere, as per the method and reasoning laid out in \citep{Wang2019, Leiberg2021}.

After the Gaussian curvature correction of surface areas $A_t(\lambda)$ and $A_e(\lambda)$, we computed $K(\lambda)$ and $S(\lambda)$ as linear combinations of the logarithms of $A_t(\lambda)$, $A_e(\lambda)$, and $T(\lambda)$:

\begin{equation}\label{K}
    K(\lambda) = \log A_t(\lambda) + \frac{1}{4} \log T(\lambda)^2 - \frac{5}{4} \log A_e(\lambda),
\end{equation}
\begin{equation}\label{S}
    S(\lambda) = \frac{3}{2} \log A_t(\lambda) - \frac{9}{4} \log T(\lambda)^2 + \frac{3}{4} \log A_e(\lambda).
\end{equation}

We normalised values of $K$ and $S$ by the lengths of the vectors $\vec{\kappa}=\{1,\frac{1}{4},-\frac{5}{4}\}$ and $\vec{\sigma}=\{\frac{3}{2},-\frac{9}{4},\frac{3}{4}\}$. That way, the transformation from traditional measures to independent measures $K$ and $S$ corresponds to a change in morphological coordinate space to an orthonormal basis of $\frac{\vec{\kappa}}{|\vec{\kappa}|}$, $\frac{\vec{\sigma}}{|\vec{\sigma}|}$, and $\frac{\vec{\iota}}{|\vec{\iota}|}$, where $\vec{\iota}=\{1,1,1\}$ is the cross product of $\vec{\kappa}$ and $\vec{\sigma}$.

\subsection{Lifespan effects on hemisphere}

\begin{figure}[h!]
\centering
\includegraphics[scale=0.9]{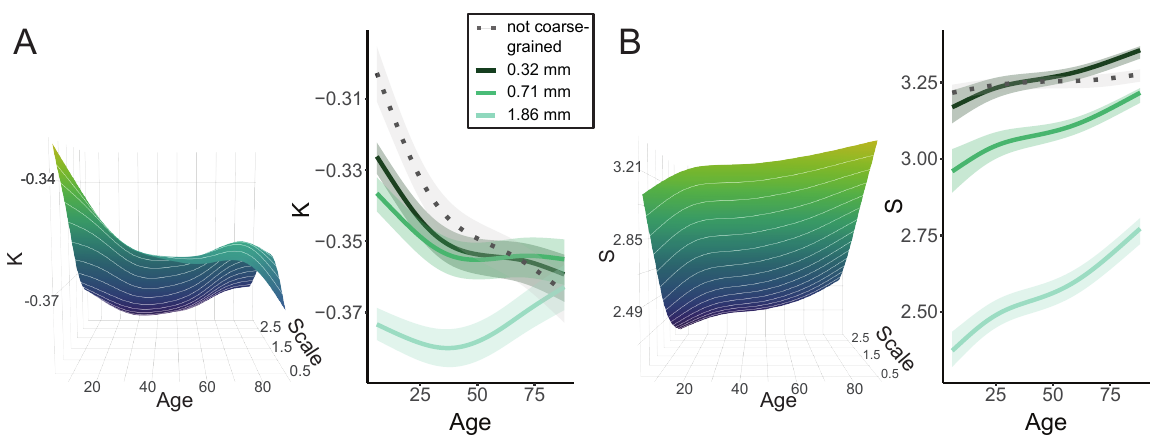}
\caption{\textbf{Lifespan effects on cortical hemispheres measured in multiscale morphometrics.} \textbf{A)}~Dimensionless metric $K$. \textbf{B)}~Dimensionless metric $S$. Sheets (left) show trajectories as functions of scales between 0.32~mm and 3~mm. Line graphs show trajectories without coarse-graining (``native scale'', dashed line) and for three scales 0.32~mm, 0.71~mm, and 1.86~mm, where lighter colour indicates a larger scale used for the coarse-graining procedure.  Shaded bands show the interquartile range of the distribution. The relative ordering of values of $S$ for the different scales results from a cortex that becomes both thicker and less gyrified as the cut-off scale increases.} \label{FigHemiKS}
\end{figure}

Figure \ref{FigHemiKS} describes lifespan effects on two independent morphometrics: the tension term~$K$ measured in smaller scales decreases rapidly until early adulthood (about 35~years), when it plateaus and then decreases further in older age (after 70~years) (Fig.~\ref{FigHemiKS}~A). In larger scales, we found an opposing trajectory that decreased only slightly until adulthood, and then increased after 60~years until older age. In the shape term~$S$, we found similar trajectories in all scales, showing a steeper increase in adolescence, which flattened in adulthood, before becoming steeper in larger scales in later life (Fig.~\ref{FigHemiKS}~B).

\clearpage
\subsection{Lifespan effects on lobes}

\begin{figure}[h!]
\centering
\includegraphics{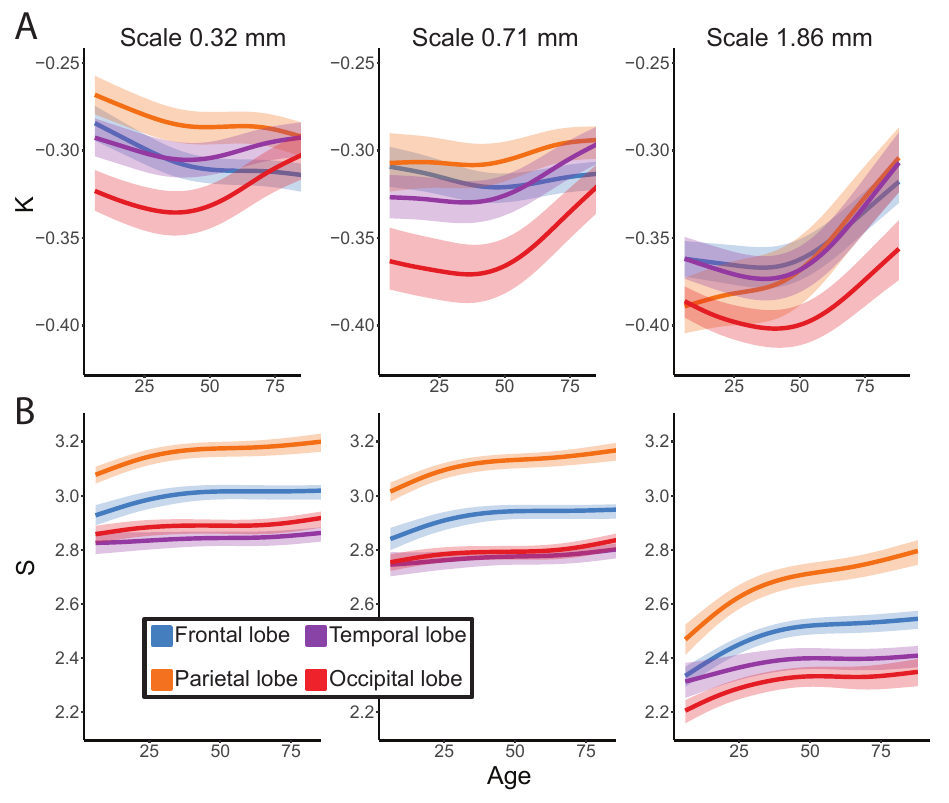}
\caption{\textbf{ Lifespan effects in main lobes measured in multiscale metrics.} Columns show trajectories in spatial scales 0.32~mm, 0.71~mm, and 1.86~mm. \textbf{A)}~Dimensionless metric $K$. \textbf{B)}~Dimensionless metric $S$. Colours indicate individual trajectories of cortical lobes. Shaded bands show the interquartile range of the distribution.} \label{FigLobeKS}
\end{figure}

In tension term $K$, regional differences in lifespan effects already become apparent at the smaller scale of 0.32~mm, where most folding details of the original surface are retained, differentiating frontal and parietal from occipital and temporal regions (Fig.~\ref{FigLobeKS}~A). In the medium scale (0.71~mm), at which smaller folds have closed, but the surface still has much gyrification, the frontal and parietal lobes saw little overall change over the lifespan, whilst temporal and occipital lobes see an increase in this morphometric and scale in later age, similar to the hemisphere trajectory in the largest scale (1.86~mm, Fig.~\ref{FigHemiKS}). In $S$ (Fig.~\ref{FigLobeKS}~B), measuring morphological complexity, we again found diverging trajectories. For example, whilst we see similar lifespan effects in the occipital lobe and temporal lobe at 0.32~mm and 0.71~mm, they have unique trajectories at 1.86~mm, demonstrating again how regional differences in lifespan become more apparent at larger scales that capture overarching morphological features.

\subsection{Interpretation}

In independent metrics $K$ and $S$, the findings at the smallest scale agreed with previously described lifespan effects without any coarse-graining \citep{DeMoraes2022}. We found that increased scales mainly affected the trajectory of the cortical surface tension, but only the offset of a measure of morphological complexity. The tension term $K$ reduces with age at small scales, but not in larger ones. This could have a biological explanation, where certain types of axons, responsible for smaller morphology features, degrade more with age. Previous literature has shown a differential effect of age in long-range fibres compared to superficial white matter \citep{Schilling2023, Schilling2023a}. However, these associations are purely speculative at this stage, and future work will have to interrogate these directly.

Regional differences in age trajectories are already apparent at small scales in the independent metrics $K$ and $S$. Generally, the trajectories of the frontal lobe were similar to those of the parietal lobe, and the occipital was similar to the temporal lobe, indicating differing lifespan effects between sensory and association cortices. $K$ remains largely unchanged at scale 0.71~mm in frontal and parietal lobes, but increases with age in occipital and temporal lobes, which may result from axonal tension in medium-length white matter tracts varying more across the lifespan in the latter two lobes. This would be contrary to some previous work on regional differences in ageing-related white matter changes, which indicated reduced integrity of prefrontal white matter \citep{Gunning-Dixon2009}. However, longitudinal studies did not find such regional differences \citep{Barrick2010}. The regional differences in $S$ show that, even though thinning is similar across regions, folding complexity changes more in the frontal and parietal regions.

\clearpage

\section{Lifespan trajectories in larger range of scales}

For cortical hemispheres, we show ``sheets'' of ageing effects measured across spatial scales in metrics cortical thickness, pial surface area, tension $K$, and shape $S$. Not that above a scale of around 5~to~6~mm, the trajectories do not differ much between scales, indiciating that at those large scales the coarse-graining algorithm has smoothed the cortical surfaces, and further incease in scale does not alter the resulting surface reconstruction much.

\begin{figure}[h!]
\centering
\includegraphics[scale=0.9]{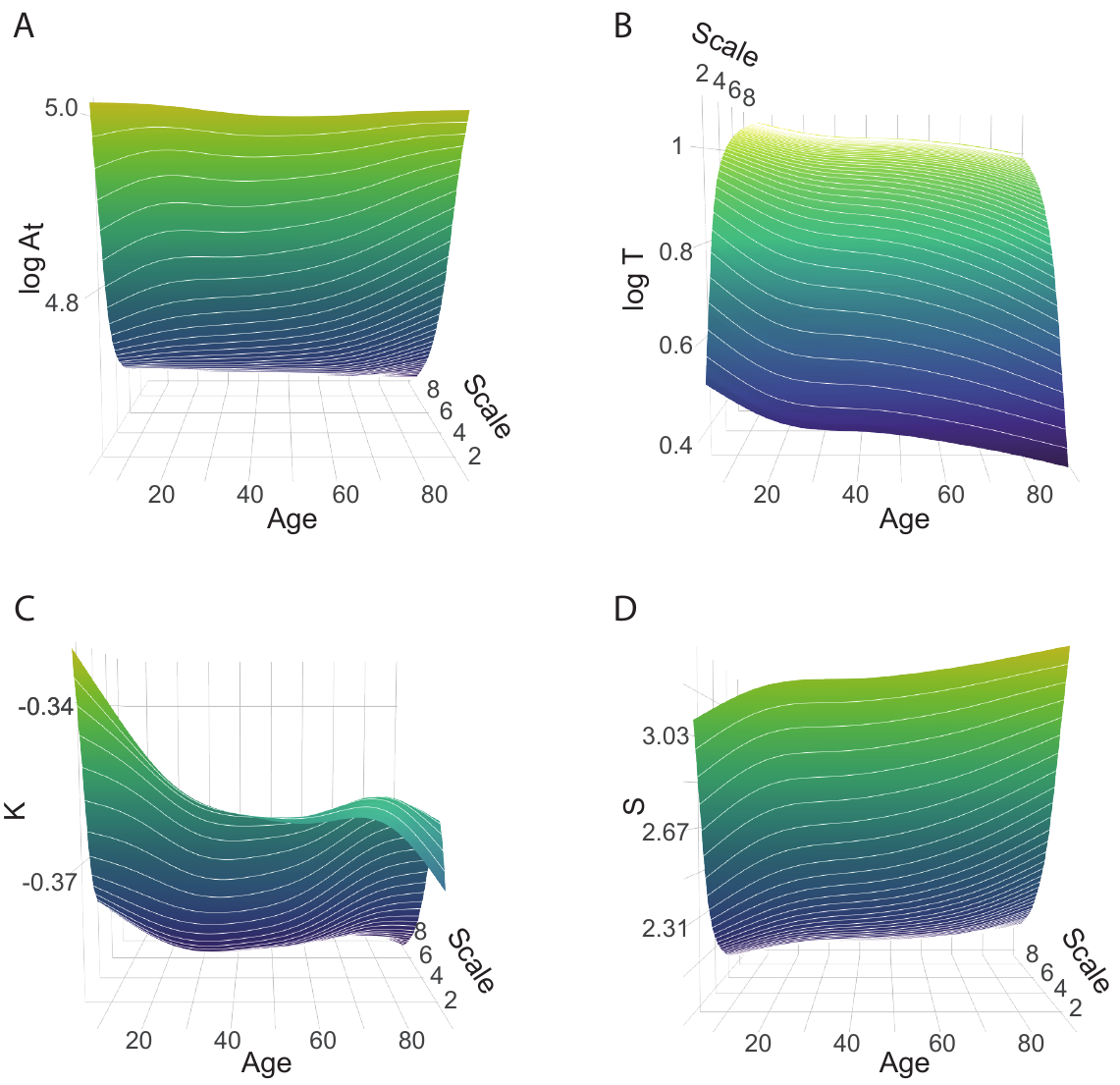}
\caption{\textbf{Lifespan trajectories in scales from 0.32~mm to 9~mm.} \textbf{A)}~Pial surface area $\log10 (A_t/mm^2)$. \textbf{B)}~Average cortical thickness $\log10 (T/mm)$. \textbf{C)}~Dimensionless metric $K$. \textbf{D)}~Dimensionless metric $S$. } \label{supplMoreScales}
\end{figure}



\end{document}